# Neutron Scattering Studies of spin excitations in hole-doped Ba$_{0.67}$K$_{0.33}$Fe$_2$As$_2$ superconductor

Chenglin Zhang[1,*], Meng Wang[2,1,*], Huiqian Luo[2], Miaoyin Wang[1], Mengshu Liu[1], Jun Zhao[1], D. L. Abernathy[3], T. A. Maier[4], Karol Marty[3], M. D. Lumsden[3], Songxue Chi [5,6], Sung Chang[5], Jose A. Rodriguez-Rivera[5,6], J. W. Lynn[5], Tao Xiang[2,7], Jiangping Hu[2,8], Pengcheng Dai[1,2,3, #]

[1]Department of Physics and Astronomy, The University of Tennessee, Knoxville, Tennessee 37996-1200, USA

[2]Beijing National Laboratory for Condensed Matter Physics, Institute of Physics, Chinese Academy of Sciences, Beijing 100190, China

[3]Neutron Scattering Science Division, Oak Ridge National Laboratory, Oak Ridge, Tennessee 37831, USA

[4]Center for Nanophase Materials Sciences and Computer Science and Mathematics Division, Oak Ridge National Laboratory, Oak Ridge, Tennessee 37831, USA

[5]NIST Center for Neutron Research, National Institute of Standards and Technology, Gaithersburg, Maryland 20899-6012, USA

[6]Department of Materials Science and Engineering, University of Maryland, College Park, Maryland 20899, USA

[7]Institue of Theoretical Physics, Chinese Academy of Sciences, P. O. Box 2735, Beijing 100190, China

[8]Department of Physics, Purdue University, West Lafayette, Indiana 47907, USA

*These authors made equal contributions to the project.

#Correspondence and requests for materials should be addressed to P.D. (pdai@utk.edu)

**We report inelastic neutron scattering experiments on single crystals of superconducting $Ba_{0.67}K_{0.33}Fe_2As_2$ ($T_c$ = 38 K). In addition to confirming the resonance previously found in powder samples, we find that spin excitations in the normal state form longitudinally elongated ellipses along the $Q_{AFM}$ direction in momentum space, consistent with density functional theory predictions. On cooling below $T_c$, while the resonance preserves its momentum anisotropy as expected, spin excitations at energies below the resonance become essentially isotropic in the in-plane momentum space and dramatically increase their correlation length. These results suggest that the superconducting gap structures in $Ba_{0.67}Ka_{0.33}Fe_2As_2$ are more complicated than those suggested from angle resolved photoemission experiments.**

High-temperature (high-$T_c$) superconductivity in iron arsenides arises from electron or hole doping of their antiferromagnetic (AF) parent compounds[1-5]. The electron pairing, as well as the long range AF order, can arise from either quasiparticle excitations between the nested hole and electron Fermi surfaces[6-11], or local magnetic moments[12-16]. In the itinerant picture, the superconducting pairing causes the opening of sign-reversed ($s^{\pm}$ wave) gaps in the respective hole and electron Fermi surfaces, as evidenced by a strong neutron spin resonance below $T_c$[17-25]. The observation of an in-plane momentum dependence of the resonance, with lengthened direction transverse to the AF wave vector $Q_{AFM}$ (Figs. 1a and 1b), in single crystals of electron-doped $BaFe_{2-x}(Co,Ni)_xAs_2$ superconductors[23-25] suggests a fully gapped $s^{\pm}$ state[26], but the predicted momentum anisotropy of the spin excitations in optimally hole-doped materials has not been observed[23]. Here we report inelastic neutron scattering experiments on single crystals of superconducting $Ba_{0.67}K_{0.33}Fe_2As_2$ ($T_c$ = 38 K). In addition to confirming the resonance[19], we find that spin excitations in the normal state form longitudinally elongated ellipses that are rotated 90° to be along the $Q_{AFM}$ direction in momentum space, consistent with the density functional theory (DFT) prediction[23]. On cooling below $T_c$, the resonance preserves its moment anisotropy as expected[23,26], but the spin excitations for energies below the resonance unexpectedly become essentially isotropic in the in-plane momentum space and dramatically increase their correlation length. These results suggest that the superconducting gap structures in $Ba_{0.67}K_{0.33}Fe_2As_2$ are more complicated than those suggested from angle resolved photoemission experiments[11].

Soon after the discovery of high-$T_c$ superconductivity in iron arsenides, band structure calculations predicted the presence of two hole-type cylindrical Fermi surfaces around the zone center ($\Gamma$ point) and electron-type Fermi surfaces near zone corners (the $M$ points, Fig. 1b)[6,7]. The unconventional electron pairing in these materials can arise from either a repulsive magnetic interaction between the hole and electron Fermi surfaces[6-9] or local AF moment exchange couplings[10,12,14], both of which necessitate a sign change in their superconducting order parameters. In the simplest picture of this so-called "$s^{\pm}$-symmetry" pairing state, nodeless superconducting gaps open everywhere on the hole and electron Fermi surfaces below $T_c$. One of the most dramatic consequences of such a state is the presence of a neutron spin resonance in the superconducting state, which occurs at the AF ordering wave vector $Q_{AFM}$ with an energy at (or slightly less than) the addition of hole and electron superconducting gap energies ($E = |\Delta(k + Q)| + |\Delta(k)|$), and a clean spin gap below the resonance[17,18]. The intensity gain of the resonance below $T_c$ is compensated by the opening the spin gap at energies below the resonance. The observed transverse momentum anisotropy of spin excitations in the electron-doped materials[23-25] favors a fully gapped $s^{\pm}$-symmetry superconductivity due to enhancement of the intraorbital, but interband, pair scattering process[26] (see also Fig. 1b and supplementary information).

*Results*

In the initial neutron scattering experiments on powder samples of hole-doped $Ba_{0.6}K_{0.4}Fe_2As_2$ ($T_c$ = 38 K), a neutron spin resonance near 14 meV was identified[19]. While the mode occurred near the AF wave vector $Q_{AFM}$ as expected, the powder nature of the experiment meant one could not obtain detailed information on the energy and wave vector dependence of the excitations[19], and therefore could not test the DFT prediction that the in-plane anisotropic momentum dependence of the spin excitations in hole-doped $Ba_{0.6}K_{0.4}Fe_2As_2$ should be rotated 90 degrees from that of the electron-doped materials (Fig. 1b)[23]. Since spin excitations can directly probe the nature the superconducting gap symmetry and phase information[18,23,26], a determination of the electron-hole symmetry in the spin dynamics of iron-arsenide superconductors is particularly crucial in view of the conflicting reports concerning the pairing symmetries by angle resolved photoemission spectroscopy (ARPES)[11,27,28], penetration depth[29], and thermo-conductivity meassurements[30-32]. Surprisingly, we find that the spin excitations at energies below the resonance in $Ba_{0.67}K_{0.33}Fe_2As_2$ ($T_c$ = 38 K, Fig. 1c) have strong sinusoidal *c*-axis modulations around $q_z = 2\pi L/c$ ($L$ = 1,3,…Fig. 1a and Fig. 2g), similar to their undoped parent compounds[33], and display clean spin gaps in the superconducting state only for energies below ~0.75 meV (Figs. 1e-1h and Fig. 2a-d). Furthermore, we discovered that the in-plane momentum dependence of the spin excitations in the normal state of $Ba_{0.67}K_{0.33}Fe_2As_2$ is elongated along the $Q_{AFM}$ direction (Figs. 4b and 4d), thus confirming the DFT prediction[23]. Although superconductivity does not change the momentum anisotropy of the resonance as expected[23,26] (Fig. 3 and Fig. 4c), the spin excitations at energies below the resonance dramatically increase the correlation length along the $Q_{AFM}$ direction and become essentially isotropic below $T_c$ (Fig. 2 and Fig. 4). Our work is not consistent with the large three-dimensional superconducting electronic gaps observed by ARPES[11,27,28]. On the other hand, given the strong gap variation over the Fermi surfaces[27,28], one might imagine that the low-energy spin excitations arise from weak nodes or small gaps not directly seen by ARPES.

Figure 1c shows the transport and magnetic properties of our single crystals of $Ba_{0.67}K_{0.33}Fe_2As_2$ which indicate $T_c$ = 38 K. Our samples were grown by using the self-flux method similar to an earlier report[34]. In previous neutron scattering and muon-spin-relaxation measurements on Sn flux grown $Ba_{1-x}K_xFe_2As_2$ single crystals[35,36], static AF order was found to phase separate from the superconducting phase due to K-chemical inhomogeneity. We have carried out systematic inductively coupled plasma atomic-emission spectroscopy analysis on our samples (size up to 15 mm*10 mm*1 mm) to confirm their chemical composition. Although our analysis also showed that K-concentrations vary slightly (up to 3%) for different batches, the superconducting properties near optimal K-doping are insensitive to such concentration variations and these samples have no static AF order coexisting with superconductivity at 2 K as shown by the neutron diffraction measurements in Fig. 1d. Independent nuclear magnetic resonance (NMR) measurements on these samples (W. P. Halperin, private communication) also confirmed the absence of the static AF order at 4 K and showed that the local magnetic field distribution as determined by the NMR linewidth is much narrower than that of the earlier K-doped BaFe$_2$As$_2$ samples[37].

To determine the energy dependence of the imaginary part of the dynamic spin susceptibility $\chi''(Q,\omega)$, we measured energy scans at the $Q_{AFM} = (1,0,0)$ and $(1,0,1)$ which correspond to spin excitations at AF wave vector transfers purely in the plane ($L = 0$) and $L = 1$, respectively (Fig. 1a), in the orthorhombic notation suitable for the parent compounds[21,33]. Figures 1e and 1g show the raw data measured on the cold neutron triple-axis spectrometer above and below $T_c$. The corresponding dynamic spin susceptibilities, $\chi''(Q,\omega)$, obtained by subtracting the background and correcting the Bose population factors (Figs. 1e and 1g), are shown in Figs. 1f and 1h for $Q = (1,0,0)$ and $(1,0,1)$, respectively. In the normal state ($T = 45$ K), $\chi''(Q,\omega)$ at both wave vectors increases linearly with increasing energy. On cooling the system to $T = 2$ K (well below $T_c$), a spin gap opens to $E = 5$ meV at $Q = (1,0,0)$, while little change of the magnetic scattering occurs at $Q = (1,0,1)$.

To confirm this conclusion, we carried out constant-energy scans at $E = 5$ meV and $E = 1.5$ meV. Figures 2a and 2b show the raw data at $E = 5$ meV across $Q = (H,0,0)$ and $(H,0,1)$, respectively. The normal state scattering shows broad peaks centered at $Q = (1,0,L)$ with $L = 0, 1$. To estimate in in-plane spin-spin correlation lengths $\xi$, we fit the scattering profile with a Gaussian on a linear background using $I = bkgd + I_0 \exp\left[-\frac{(H-H_0)^2}{2\sigma^2}\right]$, where the full width at half maximum FHWM $= 2\sqrt{2\ln 2}\sigma$ in $\text{Å}^{-1}$. Fourier transforms of the Gaussian peak in reciprocal space give normal state in-plane spin-spin correlation lengths of $\xi = \frac{2\sqrt{2\ln 2}}{\sigma} = 17 \pm 3$ and $23 \pm 4$ Å for $L = 0$ and 1, respectively[38]. Upon entering into the superconducting state, the magnetic scattering vanishes for $L = 0$, while the spin correlation length for $L = 1$ increases to $52 \pm 5$ Å (Fig. 2b). These results are consistent with Figs. 1e-1h, and confirm that the low-temperature spin gaps are strongly $L$-dependent. Similar $L$-dependence of the spin gaps have also been found in electron-doped materials[21-23]. Therefore, while superconductivity suppresses the dynamic susceptibility at $L = 0$, it dramatically increases the in-plane spin correlation length and slightly enhances $\chi''(Q,\omega)$ at $L = 1$. To see what happens at lower energies, we show in Figs. 2c and 2d constant-energy scans at $E = 0.75$, and 1.5 meV along the $(H,0,1)$ direction above and below $T_c$, respectively. While a clean spin gap is found at $E = 0.75$ meV in the superconducting state (Fig. 2c), there is clear magnetic scattering at $E = 1.5$ meV below $T_c$ (Fig. 2d).

If the opening of a spin gap as shown in Figs. 1 and 2a-d is associated with superconductivity, one should expect a dramatic reduction in magnetic scattering below $T_c$. Figure 2e shows the temperature dependence of the $E = 3$ meV scattering at the $L = 0$ signal $[Q = (1,0,0)]$ and background $[Q = (1.3,0,0)]$ positions. While the signal scattering shows a clear suppression below $T_c$ indicating the opening of a spin gap, the background scattering has no anomaly across $T_c$ and merges into the signal below 15 K. The vanishing magnetic scattering at $Q = (1,0,0)$ below $T_c$ is confirmed by $Q$-scans along the ($H$,0,0) directions (see supplementary information).  At $Q = (1,0,1)$ and $E = 3$ meV, the temperature dependence of the scattering again shows a clear suppression below $T_c$ (Fig. 2f), but in this case the background scattering at $Q = (1.4,0,1)$ does not merge into the signal at 4 K.  This is consistent with the constant-energy scans along the ($H$,0,1) directions (see supplementary information).  While the scattering shows a clear peak centered at $Q = (1,0,1)$ in the normal state, the identical scan in the superconducting state also has a peak that becomes narrower in width, indicating that the spin-spin correlation length at this energy nearly doubles from $23 \pm 3$ Å at 45 K to $40 \pm 7$ Å at 2 K.

Figure 2g shows the scattering profile in the [$H$,0,$L$] scattering plane at $E = 5$ meV and $T = 2$ K.  The magnetic signal displays a clear sinusoidal modulation along the (1,0,$L$) direction with maximum intensity at odd $L$ and no intensity at even $L$.  At $E = 3$ meV, the normal state spin excitations also exhibit a sinusoidal modulation along the $c$-axis, while the effect of superconductivity is to open spin gaps near the even $L$ positions.

Having established the behavior of the low energy spin dynamics across $T_c$, we now turn to the neutron spin resonance[19] above the spin gap energy.  In previous work on single crystals of electron-doped superconducting BaFe$_{2-x}$(Co,Ni)$_x$As$_2$ (refs. 20-25), the neutron spin resonance was found to be dispersive along the $c$-axis and occurred at significantly different energies for $L = 0$, and 1 ($\Delta E \sim$ 1-2 mV, refs. 21,23).  Furthermore, the spin excitations display larger broadening along the transverse direction with respect to $Q_{AFM}$ in momentum space without changing the spin-spin correlation lengths across $T_c$ (refs. 23-25).  To confirm the resonance in the previous powder measurements[19] and determine its dispersion along the $c$-axis for hole-doped Ba$_{0.67}$K$_{0.33}$Fe$_2$As$_2$, we carried out systematic energy scans above and below $T_c$.  Figure 3a shows the outcome at the signal [$Q = (1, 0, 2)$] and background [$Q = (1.4, 0, 2)$] positions for $T = 45$ K and 2 K. Figure 3b plots $\chi"(Q,\omega)$ at $Q = (1, 0, 2)$ across $T_c$. Inspection of Figs. 3a and 3b reveals that the effect of superconductivity is to suppress the low energy spin excitations and create a neutron spin resonance near 15 meV consistent with earlier work[19]. To test the dispersion of the resonance, we have also carried out similar measurements at $Q = (1,0,3)$. Figure 3e compares the temperature difference plots for $Q = (1,0,2)$ and $Q = (1,0,3)$, which reveals little dispersion for the neutron spin resonance in $Ba_{0.67}K_{0.33}Fe_2As_2$. This is clearly different from that of electron-doped pnictides[21,23]. Figure 3f shows the temperature dependence of the scattering at $Q = (1,0,2)$ and $E = 15$ meV. Consistent with earlier work[19], we find that the intensity of the resonance increases below $T_c$ like a superconducting order parameter. Figure 3c shows constant-energy scans at $E = 15$ meV along the ($H$,0,2) direction above and below $T_c$, and Figure 3d plots the temperature dependence of $\chi"(Q,\omega)$. These data indicate that the effect of superconductivity is to enhance the scattering at the AF wave vector without significantly changing the spin-spin correlation length.

Finally, to determine the in-plane wave-vector dependence of the spin excitations for $Ba_{0.67}K_{0.33}Fe_2As_2$, we carried out neutron time-of-flight measurements imaging the in-plane spin excitations at energies below (Figs. 4a and 4b) and at the resonance (Figs. 4c and 4d). In the normal state, the spin excitations exhibit anisotropy along the $Q_{AFM}$ direction (Figs. 4b and 4d) precisely as predicted by the DFT calculation for hole-doped materials[23]. Our calculations suggest that the longitudinal elongation in spin excitations arises from intra-orbital, inter-band scattering from $d_{yz}$ orbitals between the hole- and electron- pockets (Fig. 1b and supplementary information). In the electron-doped case, the main contribution to the spin susceptibility for $Q$ near $\mathbf{Q}_{AFM}$ = (1,0) comes from scattering processes between the blue $d_{xy}$ orbitals indicated by arrows, i.e., between the upper electron pocket and hole pocket around (1,1) in Fig. 1b.  Since the (1,1) hole pocket is quite small in the electron doped case, the nesting wave vector is offset from $\mathbf{Q}_{AFM}$ = (1,0) by a finite $\Delta Q_y$, which leads to the incommensurate peaks along the direction from (1,0) to (1,1), i.e., along the transverse direction.  In the hole-doped case, $d_{xy}$ to $d_{xy}$ scattering is still strong, but now there is also a large contribution from scattering between the $d_{yz}$ orbitals which gives rise to the incommensurate peaks along the $Q_x$ direction and therefore the longitudinal anisotropy.  The main scattering processes are again indicated by the arrows and occur between well-nested (green) regions on the outer hole pocket around $\Gamma$, and the electron pocket around (1,0).  Since the hole pocket is much larger than the electron pocket, the nesting wave-vector is offset by a finite $\Delta Q_x$ from $\mathbf{Q}_{AFM}$ = (1,0) which leads to the longitudinal anisotropy (see supplementary information for detailed calculations).

    On cooling below $T_c$, the resonance exhibits an anisotropy profile that is the same as  the spin excitations in the normal state (Figs. 4c and 4d), while the anisotropic scattering (Fig. 4b) found in the normal state for energies below the resonance becomes isotropic (Fig. 4a).  To test if the changing scattering profile for energies below the resonance is indeed associated with superconductivity, we carried out detailed temperature-dependent wave-vector measurements along the Q = [$H$, 0, $L$] ($L$ = 1) direction at $E$ = 4, 6 meV, below and above, respectively, the spin gap  of 5 meV (Fig. 1f). Figures 4e and 4g show constant-energy scans above and below $T_c$, which confirm earlier measurements at $E$ = 1.5, 3, 5 meV (Fig. 2 and supplementary information).  By fitting

the profile with a Gaussian on a linear background, we can extract the full-width-at-half-maximum (FWHM) of the scattering profile at different temperatures (see supplementary information). Figures 4f and 4h show the temperature dependence of the FWHM for $E = 4$ and 6 meV, respectively. In both cases, there is a dramatic drop in the FWHM and a corresponding increase in the spin-spin correlation length below $T_c$. Therefore, the effect of superconductivity is to induce a resonance and change the shape of the wave-vector dependent scattering from anisotropic to isotropic for excitation energies below the resonance.

*Discussion*

The novel spin excitations in hole-doped $Ba_{0.67}K_{0.33}Fe_2As_2$ we have discovered differ from the typical resonance behavior in electron-doped materials[20-25] in two important ways. First, the development of superconductivity dramatically sharpens the spin excitations for energies below the resonance and changes their dispersion from anisotropic above $T_c$ to isotropic in momentum space below $T_c$. Second, the normal spin excitations in hole-doped materials have a momentum anisotropy that is rotated 90 degrees from that of the electron-doped pnictides[23-25]. The observed elongated scattering along the longitudinal $Q_{AFM}$ direction for hole-doped superconducting $Ba_{0.67}K_{0.33}Fe_2As_2$ is consistent with the fact that incommensurate spin excitations are observed in the longitudinal direction in pure $KFe_2As_2$ (ref. 39).

In principle, spin excitations in a paramagnetic superconducting material can stem from itinerant electrons, local spin moments, or a combination of both, and can directly probe the superconducting gap symmetry[17,18]. Since there is heavy debate on whether the magnetism in iron pnictides arises from itinerant or localized electrons [6-10,40,41], we will not address this issue here but instead focus on what the temperature dependence of the spin excitations tells us about the superconducting gap structures in hole-doped materials.

From the Fermi surface nesting picture, recent DFT calculations[23] have successfully predicted that the oval shape of the normal state spin excitations in the electron-doped pnictides should rotate by 90 degrees in the hole-doped materials. This prediction can be understood from a detailed comparison of the Fermi surfaces in electron and hole doped materials (Fig. 1b), and is consistent with our observations (Figs. 4b and 4d). In the simplest $s^{\pm}$-symmetry electron pairing model[17,18], the opening of isotropic $s$-wave superconducting gaps in the hole and electron Fermi surfaces should suppress any low-energy spin excitations below the resonance. The observation of nearly zero-energy spin excitations at 2 K ($E > 0.75$ meV) and $Q_{AFM} = (1, 0, L = \text{odd})$ demonstrates that the superconducting gaps must be very small on some parts of the Fermi surfaces. These also must be linked by $Q_{AFM} = (1, 0, 1)$ and simultaneously be present on the hole and electron Fermi surfaces with sufficient phase space to account for the observed low-energy spin excitations (Figs. 1-4). The dramatic increase in the spin-spin correlation length below $T_c$ reveals that the low-energy spin excitations are also strongly affected by the opening of superconducting gaps on other parts of the Fermi surfaces. In principle, the opening of large superconducting gaps on these Fermi surfaces may reduce the scattering between the two different parts of the Fermi surfaces, which can increase the lifetime of the scattering and the spin-spin correlation length as observed in our experiments.

However, such a pure itinerant picture is inconsistent with angle resolved photoemission experiments, where the three-dimensional superconducting gaps are large in all of the observed Fermi surfaces with a minimum gap energy of 4 meV (refs. 11,27,28), a value much larger than the observed spin excitations at $L = 1$. Therefore, the only way to understand the observed low-energy spin excitations near $L = 1$ in the itinerant picture is to assume that there are significant parts of the Fermi surface (Fermi arcs), yet to be observed by photoemission experiments[11,27,28], that are essentially gapless. In this case, further calculations below $T_c$ using a random phase approximation (RPA) and the three-dimensional five-orbital tight-binding model[42] are needed to see what kind of gap structure is consistent with the observed change in the spin excitation line-width. Therefore, while our results clearly indicate that the superconducting gap structures in $Ba_{0.67}K_{0.33}Fe_2As_2$ are more complicated than those suggested by the current ARPES measurements[11,27,28], further theoretical work is necessary to understand the temperature dependence of the spin excitations.

*Methods*

Single crystals of $Ba_{0.67}K_{0.33}Fe_2As_2$ were grown the by self-flux method[43]. The resistivity and magnetic susceptibility were measured by PPMS and SQUID from Quantum design. A 6.12 mg crystal cut from a big piece used for neutron scattering shows 100% superconducting volume fraction, see the inset in Fig.1c. Many others from different batches show very similar properties. The position in reciprocal space at wave vector $\boldsymbol{Q} = (q_x, q_y, q_z)$ Å$^{-1}$ is labeled as $(H, K, L) = (q_x a/2\pi, q_y b/2\pi, q_z c/2\pi)$ reciprocal lattice units (rlu), where the tetragonal unit cell of $Ba_{0.67}K_{0.33}Fe_2As_2$ has been labeled in orthorhombic notation with lattice parameters of $a = b = 5.56$ Å, $c = 13.29$ Å (ref. 33). Our neutron scattering experiments were carried out on the HB-3 thermal neutron three-axis spectrometer at High Flux Isotope Reactor and the ARCS time-of-flight chopper spectrometers at Spallation Neutron Source, Oak Ridge National Laboratory, and on the BT-7 thermal, SPINS and MACS cold neutron triple-axis spectrometers, at the NIST Center for Neutron Research. For the HB-3, BT-7, SPINS, MACS neutron measurements, we fixed the final neutron energies at $E_f$ = 14.7 meV, 13.5 meV, 5.0 meV, and 5.0 meV, respectively. For triple-axis measurements, we co-aligned 4.5 grams of single crystals on aluminum plates. For time-of-flight measurements on ARCS, we co-aligned 60 pieces of single crystals with a total weight of 20 grams on several aluminum plates. In both cases, the in-plane and $c$-axis mosaics of aligned crystal assemblies are about 3º and 6.5º, respectively.

**Acknowledgements** This work is supported in part by the US Department of Energy, Division of Materials Science, Basic Energy Sciences, through DOE DE-FG02-05ER46202 and by the US Department of Energy, Division of Scientific User Facilities, Basic Energy Sciences. SPINS and MACS utilized facilities supported in part by the National Science Foundation under Agreement No. DMR-0454672 and No. DMR-0944772, respectively. The work at the Institute of Physics, Chinese Academy of Sciences, is supported by the Chinese Academy of Sciences. T.A.M. would like to acknowledge support from the Center for Nanophase Materials Sciences, which is sponsored at ORNL by Basic Energy Sciences, U.S. DOE.

**Author contributions** P.D. and C.L.Z. planned the experiments. C.L.Z and M.W. grew single crystals at UTK. C.L.Z. carried out thermal and cold triple-axis spectrometer measurements on HB-3, SPINS, and BT-7 with help from M.Y.W., J.Z., K.M., M.D.L., S.X.C., S.C, and J.W.L. M.W. carried out MACS and ARCS measurements with help from H.Q.L., M.S.L.,D.L.A., and J.A.R. T.A.M. carried out RPA calculations. T.A.M., T.X., and J.P.H. helped with theoretical interpretations. The paper was written by P.D., J.P.H, T.A.M. with input from all co-authors.

**Additional Information** The authors declare no competing financial interests.

**Figure 1 Schematic diagram of the reciprocal space probed, transport, and neutron scattering data on $Ba_{0.67}K_{0.33}Fe_2As_2$.** (a) Real and reciprocal space of the FeAs plane. The light and dark As atoms indicate As positions below and above the Fe-planes, respectively. (b) Fermi surfaces at $k_z=0$ calculated using the tight-binding model of Graser et al.[42] for electron and hole-doped $BaFe_2As_2$. The different colors indicate orbital weights around the Fermi surface with red = $d_{xz}$, green = $d_{yz}$, and blue = $d_{xy}$. (c) Temperature dependence of the in-plane resistivity $\rho(T)$ shows the onset of superconductivity at $T_c = 38$ K. We find no resistivity anomaly that might be associated with structural or AF phase transitions above $T_c$. The inset shows the temperature dependence of the bulk susceptibility for a 1 mT in-plane magnetic field giving $T_c = 38$ K. (d) Elastic neutron scattering along the ($H$, 0, $L$) direction with fixed values of $L = 0, 2$, and 3 at 2 K, demonstrating that there is no static AF order in our samples. (e) Energy scans at the AF signal [$Q = (1,0,0)$] and background [$Q = (1.3,0,0)$] positions from 0.5 to 8.5 meV at 45 K and 2 K. (f) $\chi''(Q,\omega)$, obtained by subtracting the background and removing the Bose population factor, clearly shows that a spin gap opens below $E \sim 5.5$ meV at 2 K. (g) Energy scans at $Q = (1,0,1)$ and background $Q = (1.3,0,1)$ positions from 0.5 to 8.5 meV at 45 K and 2 K. (h) $\chi''(Q,\omega)$, obtained using the identical method as in (f), shows quite different behavior from the results in (f). Solid lines are guides to the eye. Data in (d) are from HB-3, and those in (e-h) are from cold triple-axis SPINS. The error bars indicate one sigma throughout the paper.

**Figure 2 Wave-vector and temperature dependence of the scattering at excitation energies below the neutron spin resonance energy in $Ba_{0.67}K_{0.33}Fe_2As_2$.** (a), (b) $Q$-scans at $E = 5$ meV along the $(H,0,0)$ and $(H,0,1)$ directions above and below $T_c$. While the scattering centered at $(1,0,0)$ clearly vanishes below $T_c$, at $(1,0,1)$ it persists and sharpens. (d) $Q$-scans at $E = 1.5$ meV along the $(H,0,1)$ direction at 2 K and 45 K. The normal state peak at 45 K clearly survives superconductivity at 2 K. (c) Similar scans at $E = 0.75$ meV, where the normal state peak clearly disappears at 2 K, indicating the presence of a 0.75 meV spin gap. (e) Temperature dependence of the $E = 3$ meV scattering at the signal $[Q = (1,0,0)]$ and background $[Q = (1.3,0,0)]$ positions. The signal shows a clear suppression below $T_c$, while the background scattering goes smoothly across $T_c$. At low temperature, the signal merges into the background scattering, thus confirming the vanishing magnetic scattering. (f) Similar data at $Q_{AFM} = (1,0,1)$ and $Q = (1.4,0,1)$. While the signal also responds to $T_c$, the background scattering does not merge into the signal at low-temperature, thus indicating the continued presence of magnetic scattering. (g) Overall scattering in the $(H,0,L)$ plane at $E = 5$ meV and 2 K. The data show a clear sinusoidal modulation along the $L$-direction. Data in (a,b,d) are from SPINS, (c) from MACS, (e) from BT-7, (f) from HB-3, and (g) from MACS. The horizontal bars indicate instrumental resolutions. The slight off-centering in peak positions for different experiments from the expected $Q_{AFM} = (1,0,1)$ position is due to small sample mis-alignment problems. The scattering at $(0.6,0,L)$ with $L=0,2.5$ is of phonon or spurious origin.

**Figure 3 Energy and wave vector scans, and temperature dependence of the neutron spin resonance in $Ba_{0.67}K_{0.33}Fe_2As_2$.** (a) Energy scans from 2 to 19 meV at the signal [$Q$ = (1,0,2)] and from 2 to 17 meV at the background [$Q$ = (1.4,0,2)] positions above and below $T_c$. The background scattering has some temperature dependence below about 11 meV, probably due to the presence of phonons. This, however, does not affect the temperature dependence below 60 K. (b) $\chi''(Q,\omega)$ at $Q$ = (1,0,2) across $T_c$. (c) Constant-energy scans at $E$ = 15 meV at 45 K and 2 K. The scattering shows a well centered peak at $Q$ = (1,0,2) that increases dramatically below $T_c$, thus confirming that the mode is centered at commensurate positions. (d) The corresponding $\chi''(Q,\omega)$ at $Q$ = (1,0,2). (e) Comparison of the temperature difference (2 K minus 45 K) spectra for the neutron spin resonance at $Q$ = (1,0,2) and (1,0,3). The resonance energy is weakly $L$-dependent. (f) Temperature dependence of the scattering at the resonance energy $E$ = 15 meV and $Q$ = (1,0,2). The scattering shows a clear order-parameter-like increase below $T_c$. Data in (a-f) are from HB-3. The horizontal bar indicates instrumental resolution.

**Figure 4 The in-plane wave-vector profile of the spin excitations and temperature dependence of the scattering at energies below and near the resonance.** (a) The in-plane ($H,K$) magnetic scattering integrated from 5 meV to 10 meV at 5 K. The data were collected on the ARCS spectrometer with incident beam energy $E_i$ = 25 meV, $c$-axis along the incident beam direction. (b) Identical scan at 45 K. In both cases, the $L$-integration range is from 1.2<$L$<2.1 which corresponds to an energy integration from 5 to 10 meV. (c) The in-plane ($H,K$) magnetic scattering profile covering the resonance energy integrated between 10 meV and 18 meV at 5 K on ARCS with $E_i$ = 35 meV. (d)

Identical scan at 45 K. Here the *L*-integration range is from 1.7<*L*<3 which corresponds to an energy integration from 10 to 18 meV. The color bars indicate intensity scale and dashed circle and ellipses indicate scattering profiles. (e) Constant-energy scans for $E$ = 4 meV along the $Q$ = ($H$,0,1) direction at 2 K and 45 K with linear background subtracted. The solid lines are Gaussian fits to the data. (f) Temperature dependence of the FWHM obtained by carrying out identical scans as in (e) at different temperatures. A clear reduction in the FHWM is seen below $T_c$. (g,f) Similar data obtained for $E$ = 6 meV. Data in (a-d) are from ARCS, those in (e-h) are from MACS. See supplementary information for raw data at different temperatures.

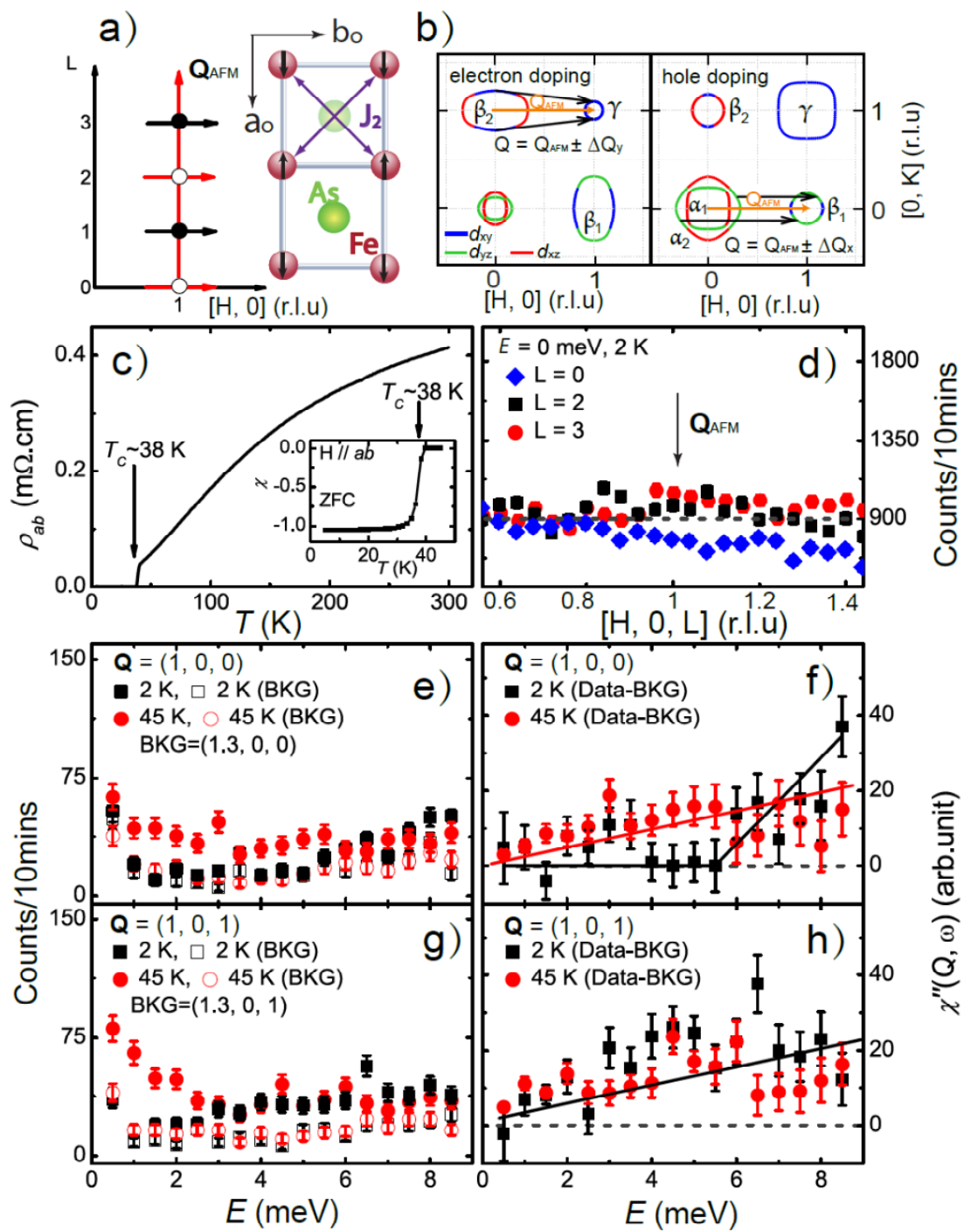

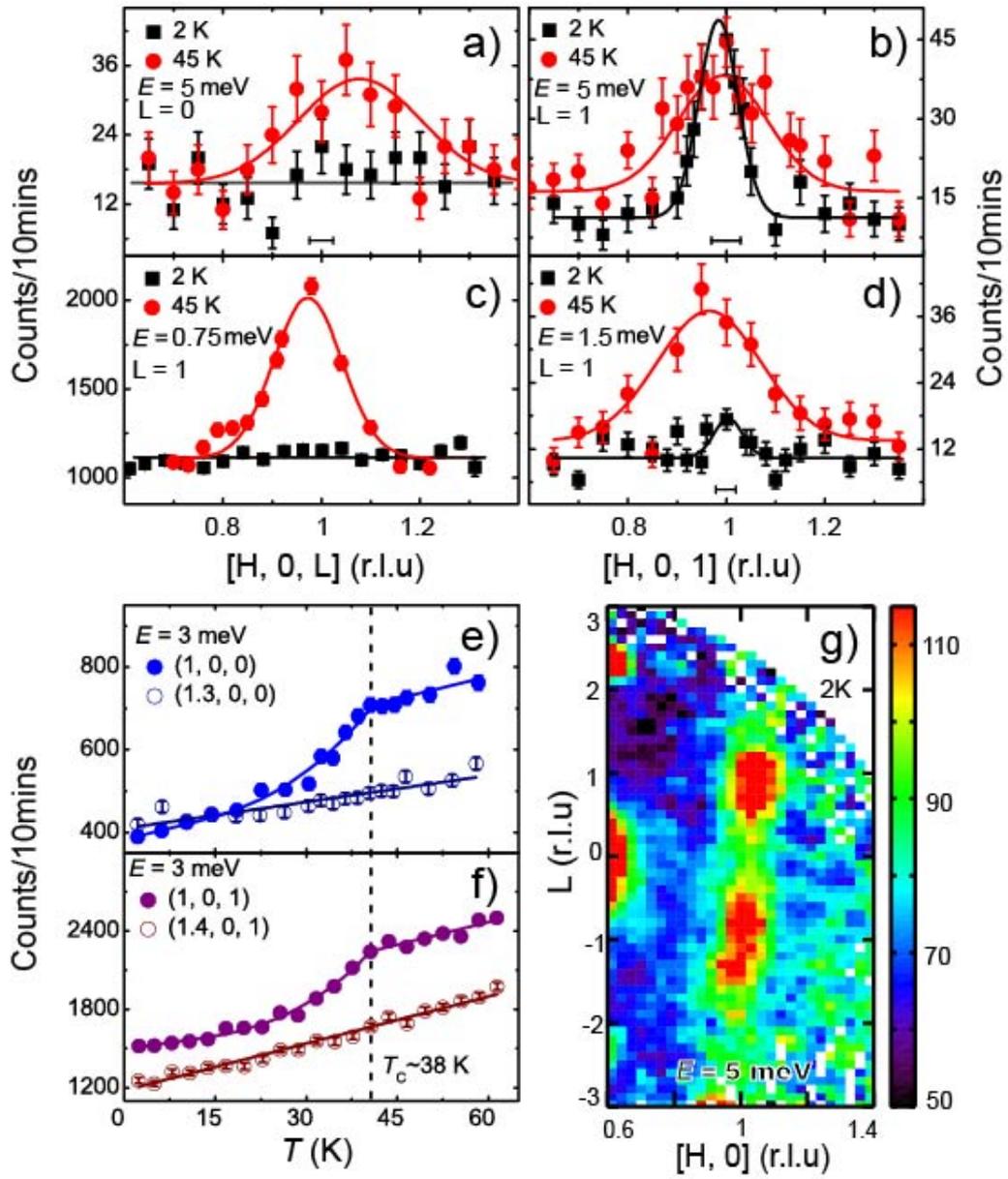

Fig. 2

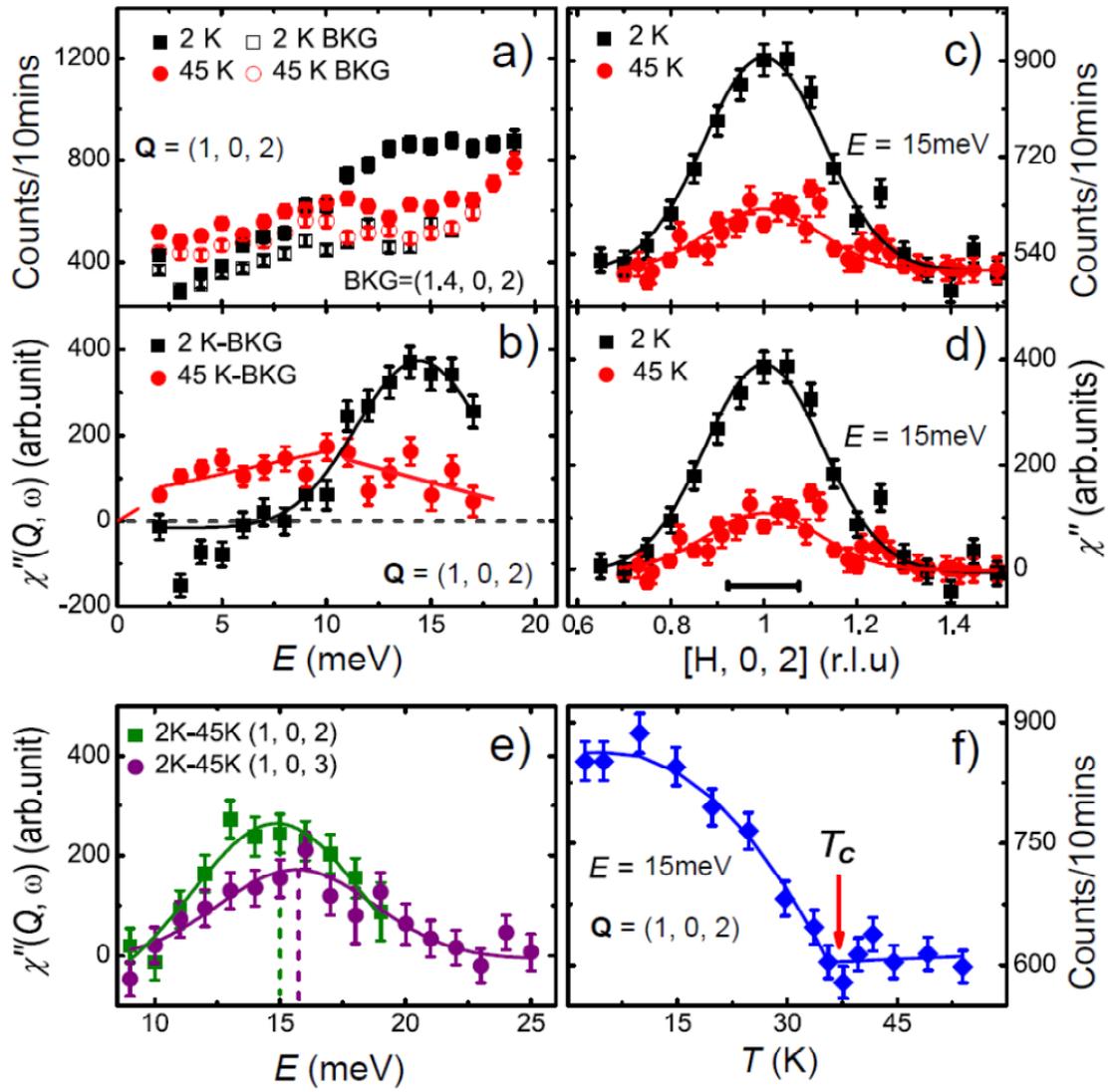

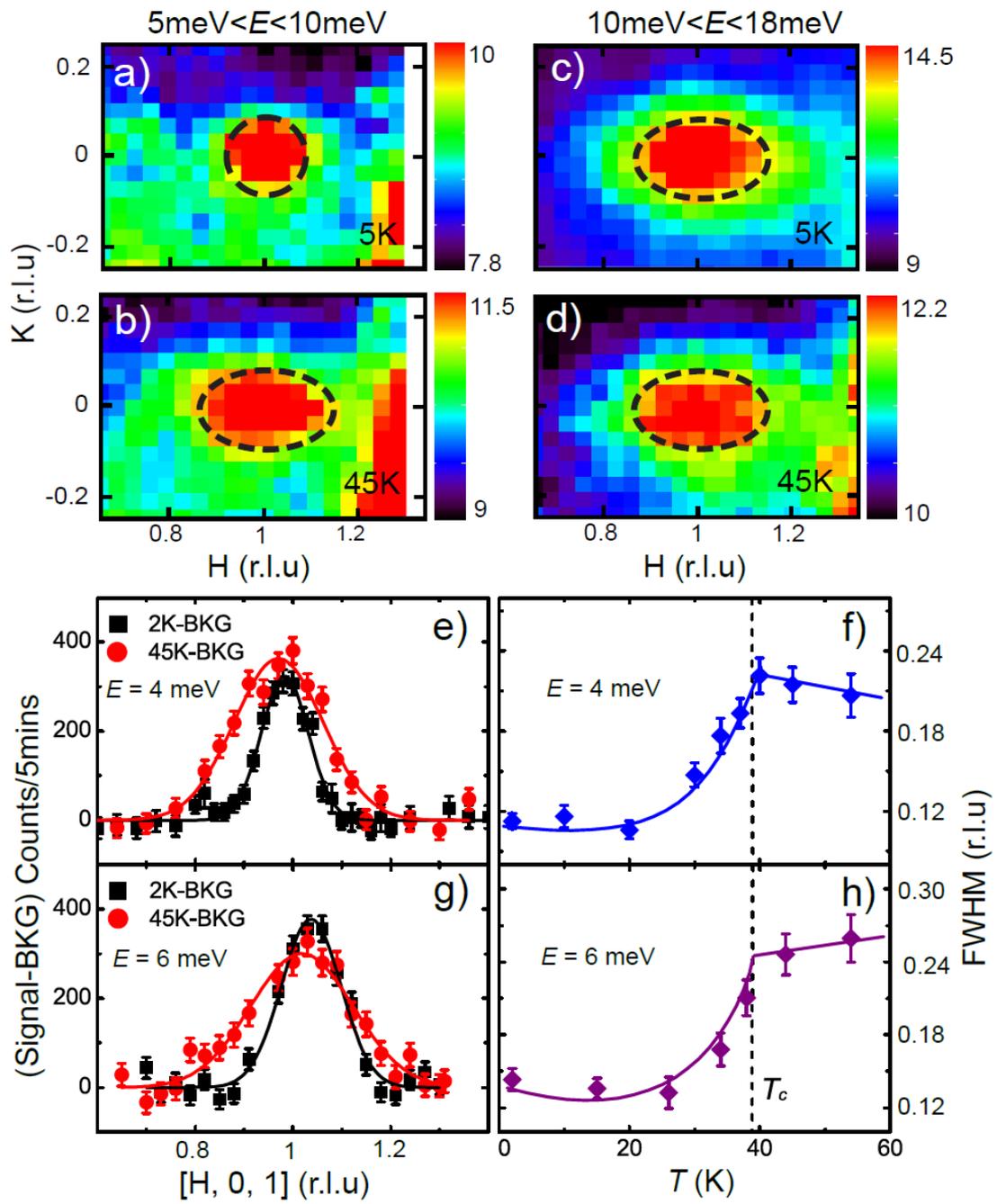

# Supplementary Information on

# Neutron Scattering Studies of spin excitations in hole-doped

# $Ba_{0.67}K_{0.33}Fe_2As_2$ superconductor

Chenglin Zhang, Meng Wang, Huiqian Luo, Miaoyin Wang, Mengshu Liu, Jun Zhao, D. L. Abernathy, T. A. Maier, Karol Marty, M. D. Lumsden, Songxue Chi, Sung Chang, Jose A. Rodriguez-Rivera, J. W. Lynn, Tao Xiang, Jiangping Hu, Pengcheng Dai

In this supplementary information, we present raw data and discuss experimental details. The instrumentation setup and description for ARCS at spallation neutron source can be found at http://www.sns.gov/instruments/SNS/ARCS/ and MACS at NIST center for neutron research is described in ref. 1. Figure SI1 shows the temperature and wave vector dependence of the scattering at $E = 3$ meV. Similar to data at E = 1.5, 4, 5, and 6 meV (Figs. 1-4), we find a clean spin gap at $L$ = even and narrowing of the $Q$-width along the $(H, 0, L)$ $L$ = odd direction. Figure SI2a shows $c$-axis modulations of the $E = 5$ meV excitations obtained on MACS. To test the spin excitation anisotropy in the in-plane moment space, we cut the ARCS data in Figs. 4a-4d along the $(H,0)$ and $(1,K)$ directions below and above $T_c$. For energy transfers below the resonance, while the scattering profile along the $(H,0)$ direction clearly narrows below $T_c$ (Fig. SI2b), it remains essentially unchanged along the $(1,K)$ direction (Fig. SI2c). At the resonance energy, we find that superconductivity has no influence on the wave vector dependence of the scattering (Figs. SI2d and SI2e). Figure SI3 summarizes the temperature dependence of the $(H,0,1)$ scattering at $E = 4$ meV obtained on MACS. Similar data at E = 6 meV are shown in Fig. SI4.

To quantitatively demonstrate the effect of electron- and hole-doping to the line shape of the spin excitations, we have carried out calculations using the random phase approximation (RPA) and the three-dimensional five-orbital tight-binding model obtained in Ref. [42] from fits of the DFT band structure for $BaFe_2As_2$. Figures **SI5** and **SI6** show intensity maps of the calculated in-plane spin susceptibility $\chi''(Q,\omega)$ at a fixed energy $E = 15$ meV for wave-vectors around $Q_{AFM}$ for a 35% hole-doped system (Fig. SI5) and an 8% electron-doped system (Fig. SI6). The calculation is based on a RPA for the three-dimensional five-orbital tight-binding model introduced in Ref. 42 for $BaFe_2As_2$. The RPA spin susceptibility $\chi(Q,\omega) = \frac{1}{2}\sum_{l_1,l_2} \chi^{RPA}_{l_1 l_1 l_2 l_2}(Q,\omega)$ is obtained from the RPA multi-orbital susceptibility matrix $\chi^{RPA}_{l_1 l_1 l_2 l_2}$, which is related to the bare (Lindhard) susceptibility matrix $\chi^0_{l_1 l_1 l_2 l_2}(Q,\omega)$ and the interaction parameters through $\chi^{RPA}(Q,\omega) = \chi^0(Q,\omega)[1 - U^s\chi^0(Q,\omega)]^{-1}$. Here $U^s$ is the interaction matrix in orbital space in the spin channel defined in Ref. 42 and contains on-site matrix-elements for the intra- and inter-orbital Coulomb repulsions $U$ and $U'$, and for the Hunds-rule coupling and pair-hopping terms $J$ and $J'$. For this calculation we have used spin-rotationally invariant parameters $U'=U-2J$ and $J' = J$ with $U = 0.8$ eV and $J = 0.2$ eV. The longitudinal and transverse shape anisotropies in the hole- and electron-doped cases, respectively, can clearly be seen and are similar to what has been observed in the experiments.

**SI References:**

1. Rodriguez, J. A. *et al.*, MACS- a new high intensity cold neutron spectrometer at NIST, *Meas. Sci. Technol*. **19**, 034023 (2008).

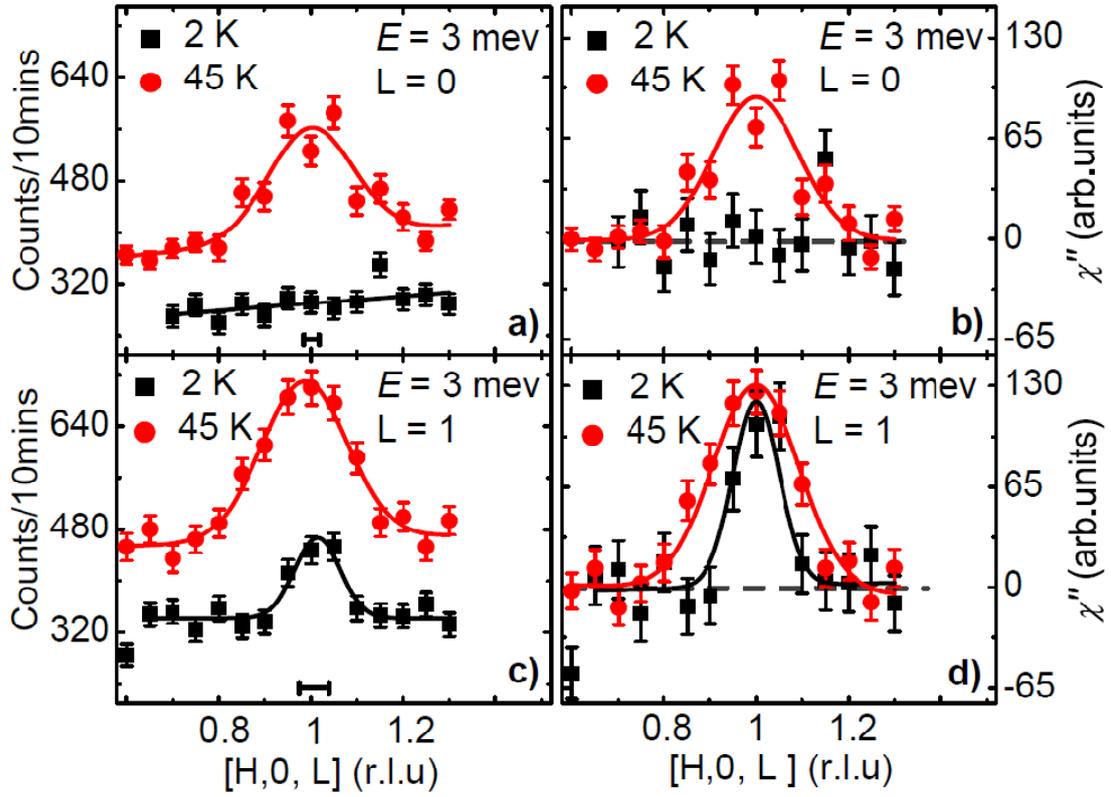

**Fig. SI1**

**Figure SI1 Temperature dependence of the wave vector scans at $E$ = 3 meV for Ba$_{0.67}$K$_{0.33}$Fe$_2$As$_2$.** (a-d) show constant energy scans through the $(H,0,L)$ $L$ = odd, even, at temperatures above and below $T_c$.

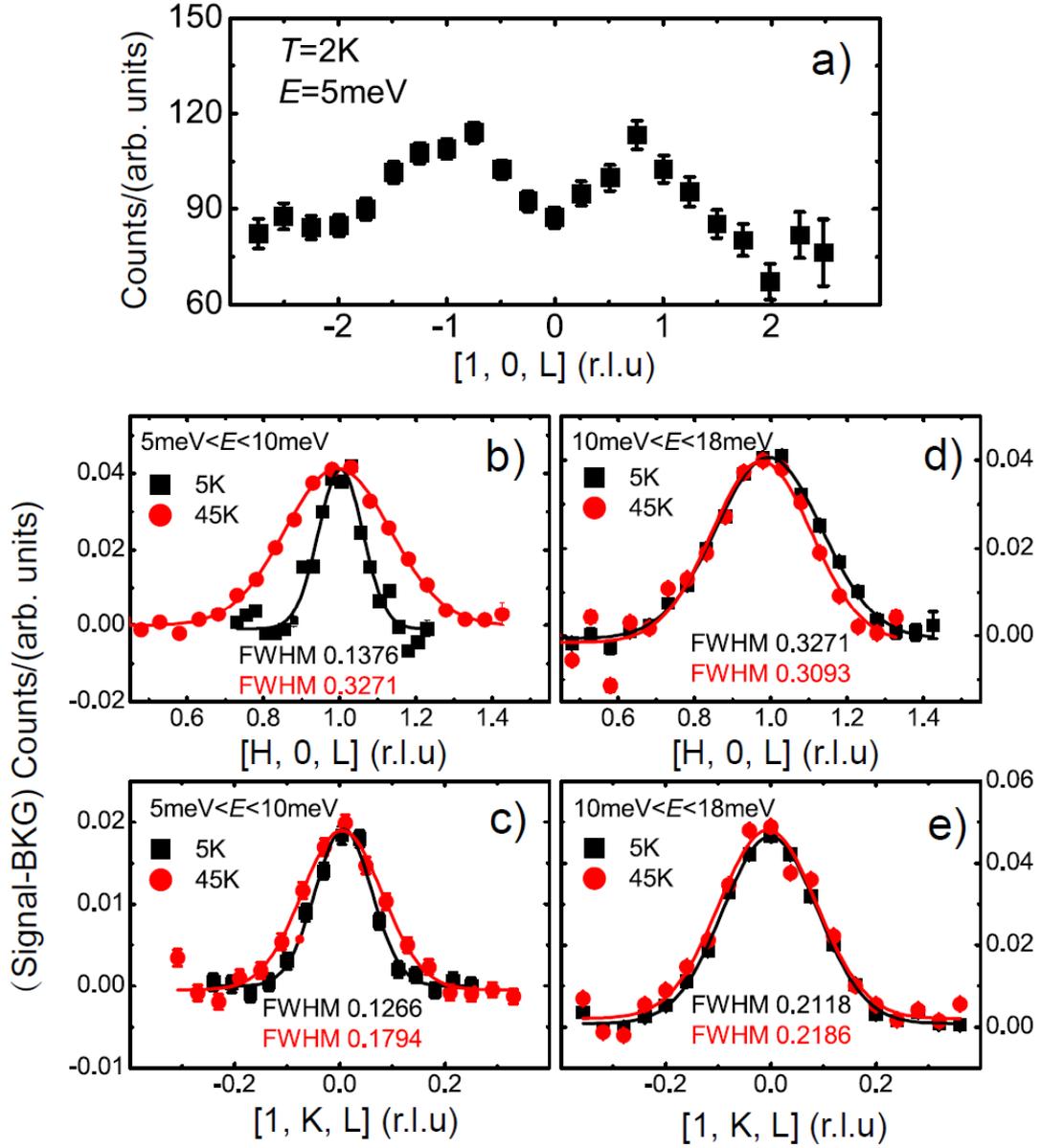

Fig. SI2

**Figure SI2 Wave vector and temperature dependence of the magnetic scattering in Ba$_{0.67}$K$_{0.33}$Fe$_2$As$_2$.** (a) $Q$-scans at $E = 5$ meV along the (1,0,$L$) directions at 2 K obtained on MACS. The integration range for $H$ is from 0.92 to 1.08 rlu. (b-d) are from ARCS. For the $H$-direction cut, the $K$-integration range is from -0.1 to 0.1 rlu. For the $K$-direction cut, the $H$-integration is from 0.85 to 1.15 rlu.

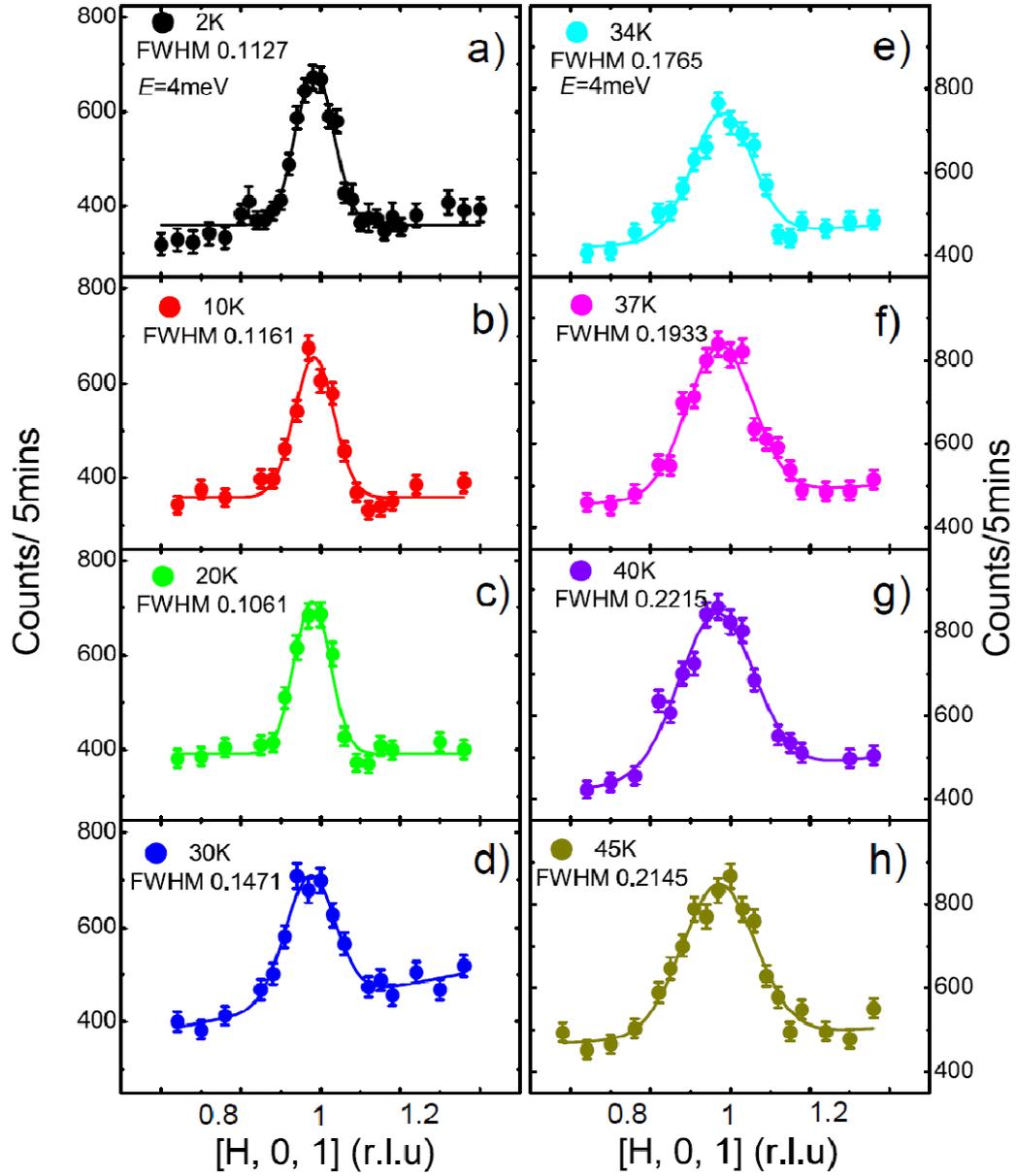

Fig. SI3

**Figure SI3 Temperature dependence of the wave vector scans at $E$ = 4 meV in $Ba_{0.67}K_{0.33}Fe_2As_2$.** (a-h) Constant energy scans at different temperatures along the ($H$,0,1) direction at E = 4 meV. The full-width-at-half maximum (FWHM) of the Gaussian fits are listed. The data are from MACS.

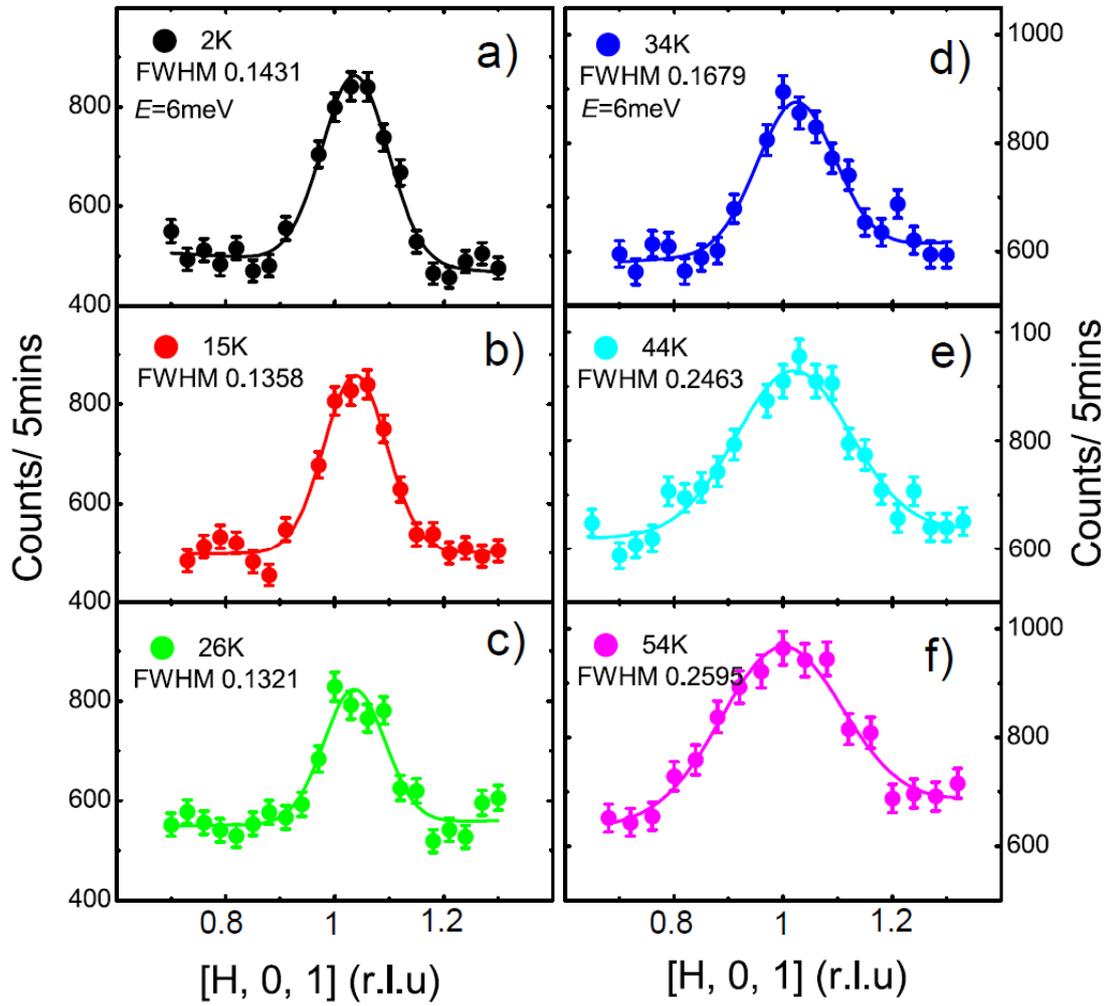

Fig. SI4

**Figure SI4 Temperature dependence of the wave vector scans at $E$ = 6 meV in Ba$_{0.67}$K$_{0.33}$Fe$_2$As$_2$.** (a-h) Constant energy scans at different temperatures along the ($H$,0,1) direction at E = 6 meV. The FWHM of the Gaussian fits are listed. The data are from MACS.

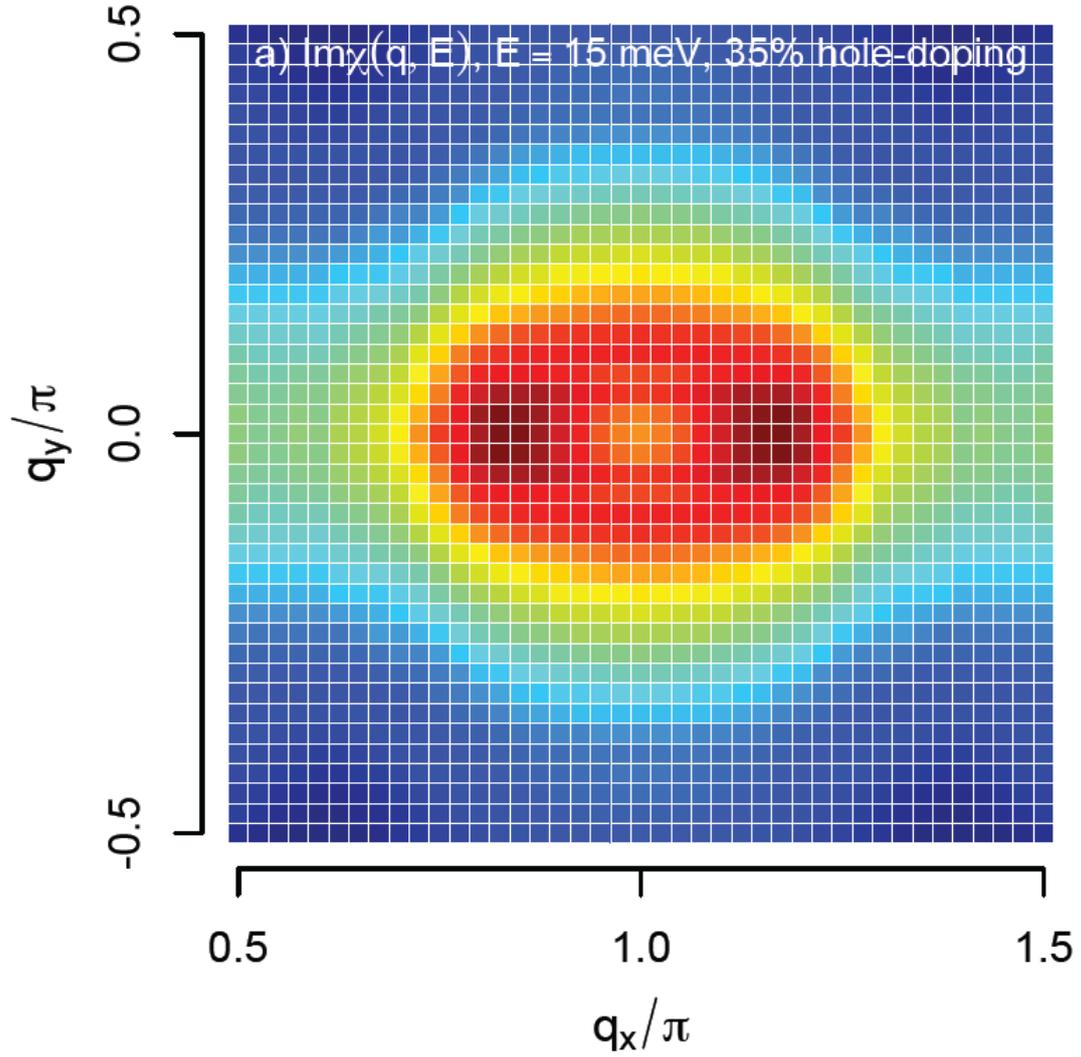

**Figure SI5 Imaginary part of the RPA spin susceptibility $\chi''(Q,\omega)$ for the hole-doped case.** The image was calculated from a three-dimensional tight-binding model for fixed energy $E = 15$ meV as a function of momentum $Q = (Q_x, Q_y, 0)$ around $Q_{AFM} = (1,0)$ for a 35% hole-doped case.

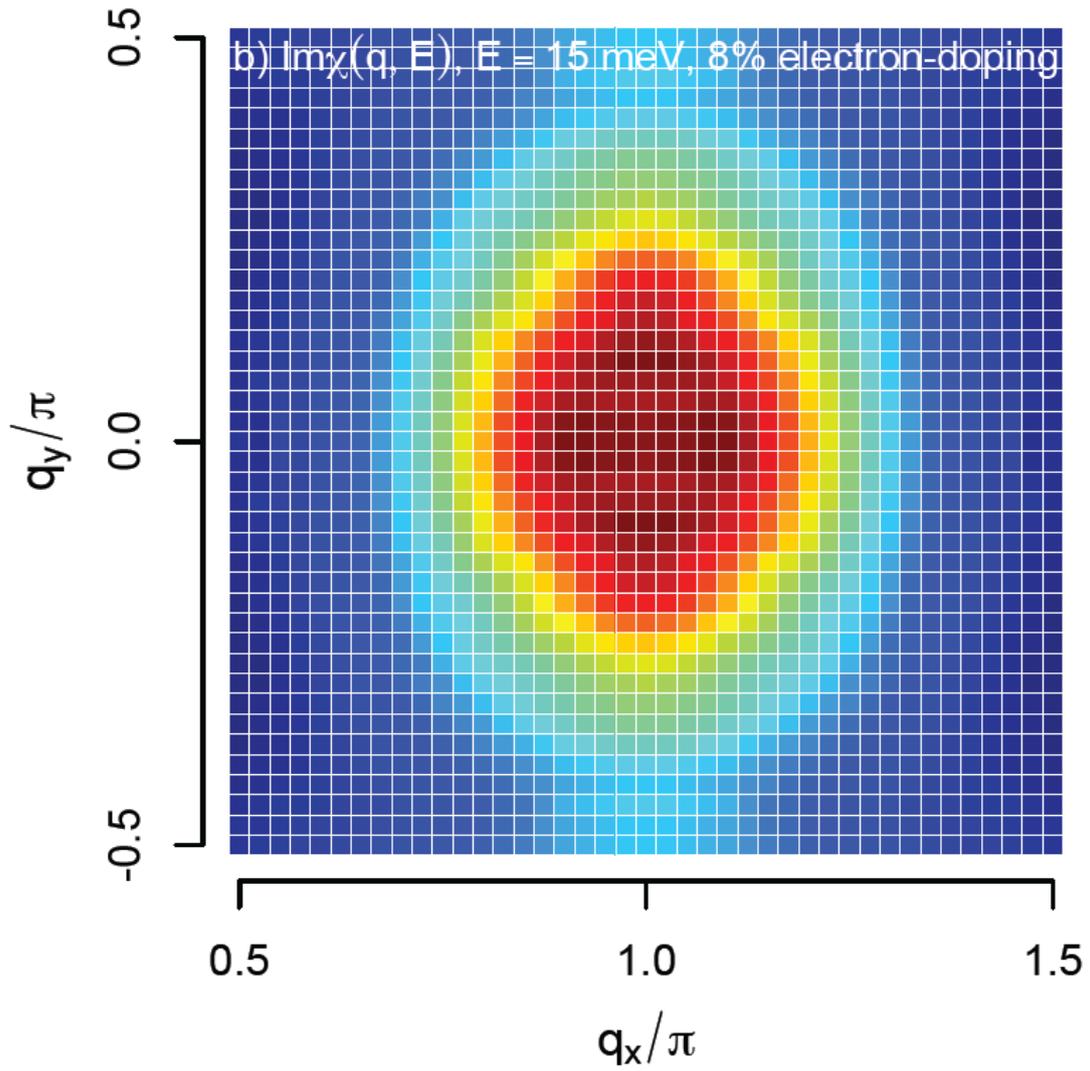

**Figure SI6 Imaginary part of the RPA spin susceptibility χ''(Q,ω) for the electron-doped case.** The image was calculated from a three-dimensional tight-binding model for fixed energy $E = 15$ meV as a function of momentum $Q = (Q_x, Q_y, 0)$ around $Q_{AFM} = (1,0)$ for an 8% electron-doped case.